\documentstyle[12pt,graphicx]{article}
\title{Black-hole horizon and metric singularity at the brane separating
two sliding superfluids}
\author{G.E. Volovik
\\Low Temperature Laboratory, Helsinki University of
Technology\\
P.O.Box 2200, FIN-02015 HUT, Finland\\
and\\
  L.D. Landau Institute for
Theoretical Physics\\  Kosygin Str. 2, 117940 Moscow, Russia}
\begin{document}
\maketitle

\abstract{An analog of black hole can be realized in the low-
temperature
laboratory. The horizon can be constructed for the `relativistic'
ripplons
(surface waves) living on the brane. The brane is represented by the
interface between two superfluid liquids,
$^3$He-A and $^3$He-B, sliding along each other without friction.
Similar
experimental arrangement has been  recently used for the observation
and
investigation of the Kelvin-Helmholtz type of instability in
superfluids
\cite{Kelvin-HelmholtzInstabilitySuperfluids}. The
shear-flow instability in superfluids is characterized by two critical
velocities. The lowest threshold measured in recent experiments
\cite{Kelvin-HelmholtzInstabilitySuperfluids} corresponds to
appearance of the ergoregion for ripplons. In the
modified geometry this will give rise to the black-hole event horizon
in the effective metric experienced by ripplons. In the region
behind the horizon, the brane vacuum is unstable due to interaction
with
the higher-dimensional world of bulk superfluids. The time of the
development of instability can be made very long at low temperature.
This
will allow us to reach and investigate the second critical velocity --
  the
proper Kelvin-Helmholtz instability threshold. The latter corresponds
to
the singularity  inside the black hole, where the determinant of the
effective metric becomes infinite.}

\maketitle
\vspace{5mm}
  PACS: 04.50.+h, 04.70.Dy, 47.20.Ft, 67.57.De
\vfill\eject
{\looseness-1
{\bf 1.~Introduction.}
The first experimental realization of two superfluid liquids sliding
along
each other \cite{Kelvin-HelmholtzInstabilitySuperfluids} gives a new
tool
for investigation of many physical phenomena related to different
areas of
physics (classical hydrodynamics, rotating Bose condensates,
cosmology,
brane physics, etc.). Here we discuss how this experimental
arrangement
can be modified in order to produce an analog of the  black-hole event
horizon and of the singularity in the effective Lorentzian metric
experienced by the collective modes (ripplons) living on the brane
(the
interface separating two different superfluid vacua,
$^3$He-A and $^3$He-B, which we further refer as the AB-brane).

}
The idea of the experiment is similar to that discussed by
Sch\"utzhold
and Unruh \cite{SchutzholdUnruh}, who suggested to use the
gravity waves on the surface of a liquid flowing in a shallow basin.
In
the long-wavelength limit
the energy spectrum of the surface modes becomes `relativistic', which
allows us to describe the propagating modes in terms of the effective
Lorentzian metric.   Here we discuss the modification of this idea to
the
case of the ripplons propagating along the membrane between two
superfluids.

{\looseness-1
There are many advantages when one uses the superfluid
liquids instead of the conventional ones: (1) The superfluids can
slide
along each other without any friction until the critical velocity is
reached, and thus all the problems related to viscosity disappear.
(2) The
superfluids represent the quantum vacua similar to that in
relativistic
quantum field theories (RQFT) (see review \cite{review}). That is why
the
quantum effects related to the vacuum in the presence of exotic metric
can be simulated. (3) The interface between two different superfluid
vacua is analogous to the brane in the modern RQFT, and one can study
the brane physics, in particular the interaction between the brane
matter
and the matter living in the higher-dimensional space oiutside the
brane.
Here, on example of the AB-brane, we show that this interaction leads
to
the vacuum instability in the AB-brane behind the event horizon.
(4) Reducing the temperature one can make the time of the development
of
the instability long enough to experimentally probe the singularity
within
the black hole (the so-called physical singularity).

}
{\bf 2.~Effective metric for modes living in the AB-brane.}
Let us consider surface waves --  ripplons -- propagating along the
AB-brane in the slab geometry shown in Fig.1.

\begin{figure}
  \centerline{\includegraphics[width=0.7\linewidth]{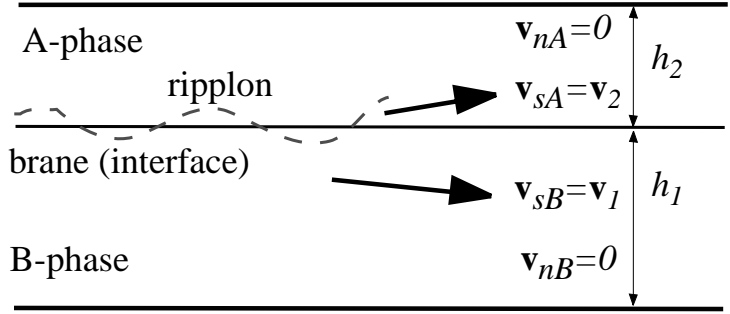}}
   \caption{The brane~-- interface between two moving superfluids,
$^3$He-A and $^3$He-B. ${\bf v}_{sA}$ and ${\bf v}_{sB}$ are the
superfluid velocities of two liquids sliding along the brane, while
the
normal components of the liquids~-- analog of the matter living
outside
the brane~-- are at rest in the frame
of container,  ${\bf v}_{nA}={\bf v}_{nB}=0$. The dashed line
demonstrates
the propagating surface wave (ripplon) which represents the matter
living
on the brane}
   \label{ABInterfaceInSlabFig}
\end{figure}

Two superfluids, $^3$He-A and $^3$He-B,
separated by the AB-brane are moving along the brane  with velocities
${\bf v}_1$ and ${\bf v}_2$ in the container frame.  The normal
components of the liquids -- the systems of quasiparticles on both
sides
of the interface -- are at rest with respect of the container walls in
equilibrium, ${\bf v}_{\rm n}=0$. The dispersion relation for
ripplons can
be obtained by modification of the equations obtained in Ref.
\cite{VolovikKH} to the slab geometry:
\vspace*{-2mm}\begin{eqnarray}
&M_1(k) (\omega- {\bf k}\cdot {\bf v}_1)^2
+ M_2(k)(\omega- {\bf k}\cdot {\bf v}_2)^2=&\nonumber\\
&=  F+k^2\sigma -   i \Gamma \omega.&
\label{GeneralSpectrum}
\end{eqnarray}
Here $\sigma$ is the surface tension of the AB-brane; $F$ is the
force stabilizing the position of the brane, in experiment
\cite{Kelvin-HelmholtzInstabilitySuperfluids} it is an applied
magnetic-field gradient;
$M_1(k)$ and
$M_2(k)$ are masses of the two liquids involved into the oscillating
motion
of the brane:
\begin{equation} M_1(k)=\frac{\rho_1}{k ~{\rm
tanh}~kh_1},~M_2(k)=\frac{\rho_2 }{k~ {\rm tanh}~kh_2}; \label{Masses}
\end{equation}
$h_1$ and $h_2$ are thicknesses of layers of two superfluids; $\rho_1$
and $\rho_2$ are mass densities of the liquids, we assume that the
temperature is low enough so that the normal fraction of each of
two superfluid liquids is small.

Finally
$\Gamma$ is the coefficient in front of the friction force
experienced
by the AB-brane when it moves with respect
to the 3D environment along the normal $\hat{\bf z}$ to the brane,
${\bf F}_{\rm fr}=-\Gamma ({\bf v}_{\rm brane}-{\bf v}_{n})$ (in the
frame of container  ${\bf v}_{n}=0$). The friction term in
Eq.(\ref{GeneralSpectrum}) containing the parameter
$\Gamma$ is the only term which couples the 2D  brane with the 3D
environment. If $\Gamma=0$, the brane subsystem becomes Galilean
invariant; the $\Gamma$-term violates  Galilean invariance in the 2D
world of the AB-brane.

In a thin slab where $kh_1\ll 1$ and $kh_2\ll 1$ one obtains
\begin{eqnarray}
&\alpha_1  (\omega- {\bf k}\cdot {\bf v}_1)^2
+ \alpha_2(\omega- {\bf k}\cdot {\bf v}_2)^2=&\nonumber\\
&= c^2k^2\left(1+ {k^2\over k_{\rm P}^2}\right) -   2i
\tilde\Gamma(k)
\omega,&
\label{ThinSlabsSpectrum}
\end{eqnarray}
where
\begin{eqnarray}
\alpha_1  = \frac{h_2\rho_1}{ h_2\rho_1+h_1\rho_2},~
\alpha_2  =1-\alpha_1= \frac{h_1\rho_2}{ h_2\rho_1+h_1\rho_2},
\label{ThinSlabsSpectrumParameters2}
\end{eqnarray}
\begin{equation}
k_{\rm P}^2= {F\over\sigma},~~~c^2=
   {Fh_1h_2\over h_2\rho_1+h_1\rho_2},
\tilde\Gamma(k) ={\Gamma\over 2}k^2 {h_1h_2\over h_2\rho_1+h_1\rho_2}
\label{ThinSlabsSpectrumParameters2}
\end{equation}
For $k\ll k_{\rm P}$ the main part of Eq. (\ref{ThinSlabsSpectrum})
can be
rewritten in the Lorentzian form
\begin{eqnarray}
g^{\mu\nu}k_\mu k_\nu= 2i\omega\tilde\Gamma(k) -c^2k^4/k_{\rm P}^2,
\label{PGwithDissipationAndNonlinear}
\\
k_\mu=(-\omega,k_x,k_y),~~~k =\sqrt{k_x^2+k_y^2},
\label{PGwithDissipationAndNonlinear2}
\end{eqnarray}
while the right-hand side of Eq.(\ref{PGwithDissipationAndNonlinear})
contains the remaining small terms violating Lorentz invariance  --
attenuation of ripplons due to the friction and their nonlinear
dispersion. Both terms come from the physics which is `trans-
Planckian'
for the ripplons.  The quantities
$k_{\rm P}$ and
$ck_{\rm P}$ play the role of the Planck momentum and Planck energy
within the brane: they determine the scales where the Lorentz
symmetry is
violated. The  Planck scales of the 2D physics in brane are actually
much
smaller than the `Planck momentum' and `Planck energy' in the 3D
superfluids outside the brane. The parameter $\Gamma$ is determined by
the physics of 3D quasiparticles scattering on the brane, and it
practically does not depend on velocities
${\bf v}_1$ and ${\bf v}_2$, which are too small for the 3D world.

At sufficiently small $k$ both non-Lorentzian terms~-- attenuation and
nonlinear dispersion on the right-hand side of
Eq.(\ref{PGwithDissipationAndNonlinear})~-- can be ignored, and the
dynamics of ripplons living on the AB-brane is described by
the following effective contravariant metric
$g^{\mu\nu}$:
\begin{equation}
\begin{array}{c}
g^{00}=-1,~~~g^{0i}=-\alpha_1 v_1^i - \alpha_2
v_2^i,\\[1mm]
g^{ij}=c^2\delta^{ij}-\alpha_1 v_1^iv_1^j - \alpha_2 v_2^i
v_2^j.\end{array}
\label{ThinSlabsRipplonContravMetric}
\end{equation}
Introducing relative velocity ${\bf U}$ and the mean velocity ${\bf
W}$
of two superfluids:
\begin{equation}
   {\bf W}=\alpha_1{\bf v}_1 +\alpha_2{\bf
v}_2,~ ~{\bf U}= {\bf v}_1 - {\bf
v}_2,
\label{NewVelocities}
\end{equation}
one obtains the following expression for the effective contravariant
metric
\begin{equation}
\begin{array}{c}
g^{00}=-1,\\ [1mm]
g^{0i}=-W^i,~~g^{ij}=c^2\delta^{ij}-W^iW^j -
\alpha_1\alpha_2 U^i U^j.\end{array}
\label{ThinSlabsRipplonContravMetric2}
\end{equation}

{\bf 3.~Horizon and singularity.}
The original KH instability
\cite{ThomsonLandauLifshitz} takes
place when the relative velocity
$U$ of the motion of the two liquids  reaches the critical value
$U_c=c/\sqrt{\alpha_1\alpha_2}$. At this velocity the
determinant of the metric tensor
\begin{equation}
    g(r)=- {1\over c^2\left(c^2-\alpha_1\alpha_2 U^2(r)\right)},
\label{ThinSlabsRipplonMetricDeterminant}
\end{equation}
   has a physical singularity: it crosses the infinite value and
changes sign. However, before
$U$ reaches
$U_c$, the system reaches the other important thresholds at which
analogs
of ergosurface and horizon in general relativity appear.  To
demonstrate
this, let us  conisder the simplest situation when velocities
${\bf U}$ and
${\bf W}$ are parallel to each other (i.e.
${\bf v}_1$ and   ${\bf v}_2$ are parallel); and these velocities
are radial and depend only on the radial coordinate $r$ along the
flow.
Then the interval of the effective 2+1 space-time in which ripplons
move
along the geodesic curves is
\begin{eqnarray}
&ds^2=&
\label{Interval1}
\end{eqnarray}
\begin{eqnarray*}
&=  {-(c^2 \!-\!
W^2(r)\!-\!\alpha_1\alpha_2 U^2(r))dt^2\! -\!2W(r)dtdr+dr^2 \over
c^2-\alpha_1\alpha_2 U^2(r)} +&\\[1mm]
&+r^2d\phi^2=&
\end{eqnarray*}
\begin{eqnarray}
&=-d\tilde t^2  {c^2 - W^2(r)-\alpha_1\alpha_2 U^2(r)\over
c^2 -
\alpha_1\alpha_2 U^2(r)} +&\nonumber\\
&+ {dr^2 \over c^2 - W^2(r)-\alpha_1\alpha_2
U^2(r)} +r^2d\phi^2,&
\label{Interval1}
\end{eqnarray}
\begin{eqnarray}
   d\tilde t =dt +  {W(r)dr\over c^2 -
W^2(r)-\alpha_1\alpha_2 U^2(r)}  ~.
\label{Interval3}
\end{eqnarray}

\begin{figure}
   \centerline{\includegraphics[width=0.7\linewidth]{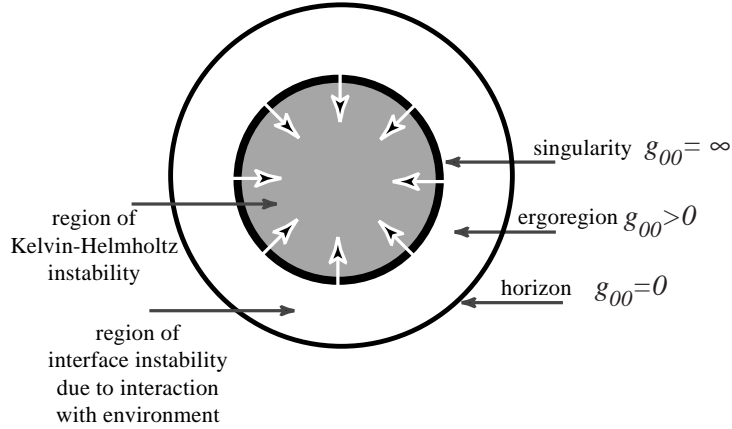}}
   \caption{Horizon and singularity in the effective metric for ripplons
on the brane (AB-interface). We assume that the A-phase is at rest, while
the B-phase is radially moving to the center as shown by arrows. }
   \label{HorizonKelvinInstabilityFig}
\end{figure}

The circle $r=r_{h}$, where  $g_{00}=0$, i.e. where
$W^2(r_{h})-\alpha_1\alpha_2 U^2(r_{h})=c^2$, marks the
`coordinate singularity' which is the black-hole horizon if the
velocity
$W$ is inward (see Fig.2). In such radial-flow geometry the
horizon also represents the ergosurface (ergoline in 2D space
dimension)
which is determined as the surface bounding the region where the
ripplon
states can have the negative energy. We call the whole region
behind the ergosurface the ergoregion.  This definition differ from
that
accepted in general relativity, but we must extend the notion of
the ergoregion to the case when the Lorentz invariance and general
covariance are violated, and the absolute reference frame appears.
At the
ergosurface, the Landau critical velocity for excitations of ripplons
is
reached. And also, as follows from Ref.
\cite{VolovikKH} (see also section {\bf 5} below), the ergoregion
coincides with the region where the brane fluctuations go unstable,
since
both real and imaginary (Fig.3) parts of the
ripplon energy spectrum cross zero at the ergosurface.

\begin{figure}
\centerline{\includegraphics[width=0.7\linewidth]{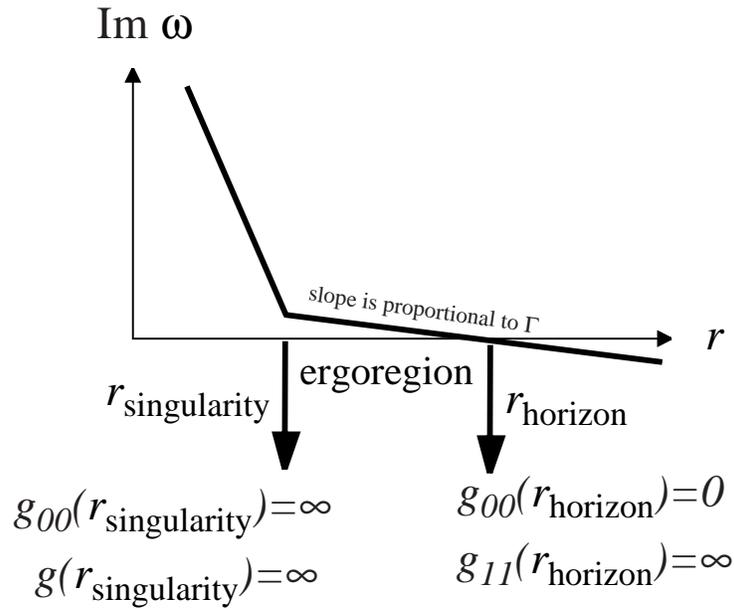}}
   \caption{Imaginary part of the ripplon spectrum due to interaction with
the environment in the higher-dimensional space. In the ergoregion the
attenuation transforms to the amplification leading to the instability of
the brane world. The time of development of this instability is long at
low $T$, where $\Gamma$ is small. On the contrary, the Kelvin-Helmholtz
instabilty behind the singularity rapidly develops and $r_{\rm
singularity}
\rightarrow 0$. }
   \label{IncrementBehindHorFig}
\end{figure}

{\bf 4.~Brane instability behind horizon.}
This means that the brane becomes unstable in the presence of the
ergoregion. This instability is caused by
the interaction of the 2D  ripplons with the 3D quasiparticles living
in
bulk superfluids on both sides of the brane \cite{VolovikKH}. The
interaction of brane with the environment, i.e. with the superfuids on
both sides of the brane, is the source  of the attenuation of the
propagating ripplons: this interaction determines the parameter
$\Gamma$
in the friction force. In the ergoregion, the imaginary part of the
spectrum of ripplons becomes positive, i.e. the attenuation tranforms
to
amplification of surface waves with negative
$\omega$ (see Figure 3 where the imaginary
part of the spectrum crosses zero with the slope proportional to
$\Gamma$). Since the instability of the interface with respect to
exponentially growing surface fluctuations develops in the presence of
the shear flow, this instability results in the formation of vortices
observed in experiment
\cite{Kelvin-HelmholtzInstabilitySuperfluids}.

In
$^3$He experiments
\cite{Kelvin-HelmholtzInstabilitySuperfluids} with shear flow along
the
AB-interface, one has
$kh_1\gg 1$ and $kh_2\gg 1$. Thus the relativistic description is
not applicable. Also, in the rotating cryostat the superfluids flow
in the
azimuthal direction instead of the radial. That is why there was no
horizon in the experiment. However, the notion of the ergosurface and
of
the ergoregion behind the ergosurface, where the ripplon energy
becomes
negative in the container frame
\cite{VolovikKH}, is applicable. The instability of the brane inside
the
ergoregion leads to formation of vortices in the vortex-free
$^3$He-B, which were
detected using NMR technique with a single vortex resolution.  The
observed threshold velocity for the vortex formation exactly
corresponds
to the appearance  of the ergosurface  (ergoline) in the container
\cite{Kelvin-HelmholtzInstabilitySuperfluids,VolovikKH}.

There are thus two ingredients which cause the vacuum
instability in the ergoregion: (i) existence of the absolute reference
frame of the environment outside the brane; (ii)  the interaction
of the brane with this environment ($\Gamma\neq 0$) which violates
Galilean (or Lorentz) invariance within the brane. They lead to
attenuation of the ripplon in the region outside the horizon. Behind
the
horizon this attenuation transforms to amplification which
destabilizes
the vacuum there. This mechanism may have an important consequence for
the astronomical black hole. If there is any intrinsic attenuation of,
say, photons (either due to superluminal dispersion, or due to the
interaction with the higher-dimensional environment),  this may lead
to
the catastrophical decay of the black hole due to instability behind
the
horizon, which we discuss in the Section {\bf 5}.

Let us estimate the time of development of such instability, first
in the artificial black hole within the AB-brane and then in the
astronomical black hole.  According to Kopnin
\cite{KopninInterface} the parameter of the friction force
experienced by
the AB-brane due to Andreev scattering of quasiparticles living in the
bulk superfluid on the A-phase side of the brane is
$\Gamma
\sim T^3 m^*/ \hbar^3c_\perp  c_\parallel$ at
$T\ll T_c$. Here
$T$ is the temperature in $^3$He-A; $m^*$ is the quasiparticle mass
in the
Fermi liquid;
$c_\perp$ and   $c_\parallel$ are the `speeds of light' for
3D quasiparticles living in anisotropic $^3$He-A (these speeds are
much
larger than the typical `speed of light' $c$ of quasiparticles
(ripplons)
living on 2D brane); $T_c$ is the superfluid transition temperature,
which
also marks the 3D Planck energy scale. Assuming  the most pessimistic
scenario in which the instability is caused mainly by the exponential
growth of ripplons with the `Planck' wave number $k_{\rm P}$, one
obtains the following estimation for the time of the development of
the
instability in the ergoregion far enough from the horizon: $\tau \sim
1/\tilde\Gamma(k_{\rm P})\sim 10 (T_c/T)^3$ sec. Thus at low $T$ the
state
with the horizon can live for a long time  (minutes or
even hours), and this life-time of the horizon can be made even
longer if
the threshold is only slightly exceeded.

This gives the unique possibility to study the horizon, the region
behind
the horizon; and the physical singularity, where the determinant of
the
metric is singular, can be also easily constructed and investigated.

At lower temperature $T<m^*c_\perp^2$ the temperature dependence
of
$\Gamma$ changes:
$\Gamma
\sim T^4  / \hbar^3c_\perp^3  c_\parallel$ \cite{review}, and at very
low
$T$ it becomes temperature independent: $\Gamma
\sim \hbar k^4$ which corresponds to the dynamical Casimir force
acting on
the 2D brane moving in the 3D vacuum.
Such intrinsic attenuation of ripplons transforms to the
amplification of the ripplon modes in the ergoregion, which leads to
instability of the brane vacuum behind the horizon even at $T=0$.

{\bf 5.~Instability of the black hole behind horizon?}
Now let us suppose that the same situation takes place in our (brane)
world, i.e. the modes of our world (photons, or gravitons, or
fermionic
particles) have finite life-time due to interaction with, say, the
extra-dimensional environment. Then this will lead to the instability
of
vacuum behind the horizon of the  astronomical black holes. This can
be
considered using the equation (\ref{PGwithDissipationAndNonlinear})
which
incorporates both terms violating the Lorentz invariance at high
energy:
the superluminal upturn of the spectrum which leads to decay of
particles, and the intrinsic broadening of the particle
spectrum characterized by
$\tilde\Gamma(k)$. Following the analogy, we can write the intrinsic
width
as a power law $\tilde\Gamma(k)
\sim \mu (ck/\mu)^n$, where $\mu$ is the energy scale which is well
above
the Planck scale
$E_{\rm P}$ of our brane world, $\mu\gg E_{\rm P}$; and
$n=6$ if the analogy is exact.

We shall use the Painlev\'e-Gullstrand
metric, which together with the superluminal dispersion of the
particle spectrum allows us to consider the region behind the horizon:
\begin{equation}
\begin{array}{c}
g^{00}=-1,~~g^{0i}=-W^i,\\[1mm]
g^{ij}=c^2\delta^{ij}-W^iW^j,~~{\bf
W}=-\hat{\bf r}\sqrt{ {2GM\over r}}.\end{array}
\label{PdGMetric}
\end{equation}
Here $G$ is Newton constant and $M$ is the mass of the black hole.
This metric coincides with the 3D generalization of the metric of
ripplons on AB-brane in Eq. (\ref{Interval1}) in case when ${\bf
v}_1={\bf
v}_2$. Equation (\ref{Interval1}) gives the following dispersion
relation for particles living in the brane:
\begin{equation}
    (\omega- {\bf k}\cdot {\bf W})^2=c^2k^2 -2i\omega\tilde\Gamma(k)
+c^2k^4/k_{\rm P}^2,
\label{PGwithDissipation}
\end{equation}
or
\begin{eqnarray}
     &\omega({\bf k})= {\bf k}\cdot {\bf W}-i\tilde\Gamma(k)
\pm&\nonumber\\
&\pm\sqrt{c^2k^2+c^2k^4/k_0^2-\tilde\Gamma^2(k) -2i{\bf k}\cdot {\bf
W}
\tilde\Gamma(k)} .&
\label{EnergySpectrum}
\end{eqnarray}
We are interested in the imaginary part of the spectrum. For small
$\tilde\Gamma(k)\ll ck$, the imaginary part of the energy spectrum is:

\noindent
\begin{equation}
\begin{array}{c}
     {\rm Im}~\omega({\bf k})=  -i\tilde\Gamma(k) \left(1\pm
     {{\bf k}\cdot {\bf W}\over E(k)} \right),\\[3mm]
E^2(k)=c^2k^2+c^2k^4/k_{\rm P}^2.\end{array}
\label{EnergySpectrumImPart}
\end{equation}
   Behind the
horizon, where
$W>c$, the imaginary part becomes positive for
$|{\bf k}\cdot {\bf W}|>E(k)$ (or $k^2 <k_{\rm P}^2/(W^2/c^2-1)$),
i.e.
attenuation transforms to amplification of waves with these ${\bf k}$.
This demonstrates the instability of the vacuum with respect to
exponentially growing electromagnetic or other fluctuations in the
ergoregion. Such an instability is absent when $\tilde\Gamma=0$, i.e.
if
there is no interaction with the trans-Planckian or
extra-dimensional world(s).

The time of the
development of instability within the conventional black hole is
determined by the region far from the horizon,  where the relevant
$k\sim k_{\rm P}$. Thus  $\tau \sim
1/\tilde\Gamma(k_{\rm P})
\sim   \mu^{n-1}/E_{\rm P}^n$.   If $\mu$ is of
the same order as the brane Planck scale, the time of
development of instability is
determined by the Planck time. That is why the astronomical black
hole can
exist only if $\mu
\gg E_{\rm P}$, which takes place when the 3D and 2D Planck scales are
essentially different, as it happens in case of the AB-brane.
The black hole decay due to the quantum process of Hawking evaporation
corresponds to $\mu=M$, where $M$ is a black hole mass, and $n=4$.

{\bf 6.~Conclusion.}
In conclusion, the AB-brane~-- the interface between the two sliding
superfluids~-- can be used to construct the artificial black hole with
the ergosurface, horizon and physical singularity. Using the AB-brane
one can also simulate the interaction of particles living on the
brane with that living in the higher-dimensional space outside the
brane.
This interaction leads to the decay of the brane vacuum in the
region behind the horizon. This mechanism can be crucial for the
astronomical black holes, if this analogy is applicable. If the matter
fields in the brane are properly coupled to, say, gravitons in the
bulk,
this may lead to the fast collapse of the black hole.

I thank  V.\,B.\,Eltsov,  M.\,Krusius, R.\,Sch\"utzhold and
W.\,G.\,Unruh for
fruitful discussions. This work was supported by ESF COSLAB Programme
and
by the Russian Foundations for Fundamental Research.

\vspace*{2mm}
\pagebreak[0]

\end{document}